\documentclass[hidelinks, colorlinks=true, allcolors=linkcolor]{eptcs}
\usepackage{enumitem}

\ifpdf
  \usepackage{underscore}         %
  \usepackage[T1]{fontenc}        %
\else
  \usepackage{breakurl}           %
\fi

\newcommand{\scasp}{{s(CASP)}\xspace}

\newcommand{\clp}{{(C)LP}\xspace}
\newcommand{\asp}{{ASP}\xspace}
\newcommand{\plai}{PLAI\xspace}

\newcommand{\csem}[1]{\ensuremath{\llbracket P \rrbracket_{#1}}}
\newcommand{\csemantics}{\csem{Q}}

\newcommand{\callpath}{\ensuremath{\mathcal{P}}\xspace} 
\newcommand{\Pt}[1]{\ensuremath{\mathcal{P}_{#1}}\xspace}
\newcommand{\csub}{\ensuremath{\theta}\xspace} 
\newcommand{\C}[1]{\ensuremath{\mathcal{C}_{#1}}}
\newcommand{\cstore}{\ensuremath{\mathcal{C}}\xspace}
\newcommand{\cstate}{\ensuremath{\Sigma}\xspace}
\newcommand{\localcstate}{\ensuremath{\sigma}\xspace}
\newcommand{\cstatetuple}[3]{\ensuremath{\langle #1,#2,#3 \rangle}\xspace}

\newcommand{\Tau}{\mathrm{T}}

\newcommand{\emptystate}{\ensuremath{\epsilon_{s}}}

\newcommand{\trace}{\texttt{Tr}\xspace}

\newcommand{\gandtree}{generalized {\sc and} tree\xspace}

\newcommand{\gandtrees}{generalized {\sc and} trees\xspace}
\newcommand{\Gandtrees}{Generalized {\sc and} trees\xspace}
\newcommand{\andtree}{{\sc and} tree\xspace}
\newcommand{\andtrees}{{\sc and} trees\xspace}

\newcommand{\literal}{\ensuremath{L}}

\newcommand{\head}{\ensuremath{H}\xspace}
\newcommand{\proj}{\ensuremath{\lambda^{pr}}\xspace}

\newcommand{\entry}{\ensuremath{\lambda^{en}}\xspace}
\renewcommand{\succ}{\ensuremath{\lambda^s}\xspace}
\newcommand{\exit}{\ensuremath{\lambda^{ex}}\xspace}
\newcommand{\builtin}{\ensuremath{\texttt{builtin}^{\#}}\xspace}
\newcommand{\asub}[1]{\ensuremath{\lambda_{#1}}}

\newcommand{\augment}{\ensuremath{\texttt{augment}^{\#}}\xspace}

\newcommand{\aproject}{\ensuremath{\texttt{proj}^{\sharp}}}
\newcommand{\project}{\texttt{proj}}
\newcommand{\extend}{\ensuremath{\texttt{extend}^{\#}}\xspace}
\newcommand{\calltoentry}{\ensuremath{\texttt{entryProc}^{\#}}\xspace}
\newcommand{\exittoprime}{\ensuremath{\texttt{exitProc}^{\#}}\xspace}

\newcommand{\entrytoexit}{\texttt{entryToExit}\xspace}

\newcommand{\amgu}{\texttt{asolve}\xspace}

\newcommand{\isloop}{\ensuremath{\texttt{loop}}\xspace}
\newcommand{\positive}{\texttt{positive}\xspace}
\newcommand{\even}{\texttt{even}\xspace}
\newcommand{\odd}{\texttt{odd}\xspace}
\newcommand{\noloop}{\texttt{noloop}\xspace}

\renewcommand{\emph}     [1]{\textit{#1}}
\renewcommand{\emptyset}    {\varnothing}

\newcommand*{\seq}      [1]{{\vec{#1}}}

\newcommand{\pred} [2]{\ensuremath{#1~\texttt{:-}~#2}}
\newcommand{\vars} [1]{\ensuremath{\mathsf{vars}(#1)}}

\newcommand{\pdesc}[1]{\ensuremath{A_{#1}}\xspace}
\newcommand{\qdesc}[1]{\ensuremath{B_{#1}}\xspace}
\newcommand{\tuple}[1]{\ensuremath{\langle #1 \rangle}\xspace}
\newcommand{\acall}[2][\lambda]{\ensuremath{#1^c_{#2}}\xspace}
\newcommand{\asucc}[2][\lambda]{\ensuremath{#1^s_{#2}}\xspace}
\newcommand{\aprime}[2][\lambda]{\ensuremath{#1^p_{#2}}\xspace}

\newcommand{\ccall}[2][\theta]{\ensuremath{#1^c_{#2}}\xspace}
\newcommand{\csucc}[2][\theta]{\ensuremath{#1^s_{#2}}\xspace}
\newcommand{\G}{\ensuremath{\mathcal{G}}\xspace}
\newcommand{\Q}{\ensuremath{\mathcal{Q}}\xspace}
\newcommand{\map}[3]{\ensuremath{\tuple{#1,#2} \mapsto #3}\xspace}

\NewDocumentCommand{\edge}{O{k} O{i} m m m m}{%
  \ensuremath{\tuple{#3,#4} \to_{#1,#2} \tuple{#5,#6}}\xspace
}
\NewDocumentCommand{\sedge}{O{k} O{i} m m}{%
  \ensuremath{ #3 \to_{#1,#2} #4 }\xspace
}

\newcommand{\fmap}[3]{\ensuremath{ \tuple{#1,#2} \overset{fx}{\mapsto} #3 }\xspace}
\newcommand{\sfmap}[2]{\ensuremath{ #1 \overset{fx}{\mapsto} #2 }\xspace}

\newcommand{\mathformula}[1]{\ensuremath{\mathsf{#1}}}
\newcommand{\del}[2]{\ensuremath{\mathformula{del}(#1,#2)}}
\newcommand{\update}[2]{{\ensuremath{\mathformula{update}(#1,#2)}}}

\newcommand{\prog}{\ensuremath{P}\xspace}
\newcommand{\aqueries}
  {\ensuremath{\Q^\sharp}}
\newcommand{\solve}
  {\ensuremath{\mathit{solve}}}
\newcommand{\eterms}{\ensuremath{\mathit{eterms}}\xspace}

\newtheorem{example}{Example}  
\newtheorem{definition}{Definition}
\newif\ifdeftriangle
\deftriangletrue   %

\makeatletter
\def\enddefinition{%
  \ifdeftriangle\hfill$\triangleleft$\fi
  \endtrivlist\@endpefalse}
\makeatother

\usepackage{ amssymb }
\usepackage{ amsmath }
\usepackage{mathtools}
\usepackage{stmaryrd}
\usepackage{algorithm}
\usepackage{wrapfig}
\usepackage[noend]{algpseudocode}
\usepackage{multicol}
\usepackage{multirow}

\usepackage{tikz}
\usetikzlibrary{shapes.geometric}
\usetikzlibrary{matrix}
\usetikzlibrary{shapes.gates.logic.US}
\usetikzlibrary{circuits}
\usetikzlibrary{positioning,shapes,arrows,fit,shadows,calc,matrix}
\usetikzlibrary{backgrounds}

\usepackage{xspace}
\usepackage{booktabs}
\usepackage{xcolor}
\definecolor{linkcolor}{HTML}{A93C93}
\usepackage{url}

\usepackage{listings}
\usepackage{adjustbox}
\usepackage{relsize}
\definecolor{PrologPredicate}{RGB}{0,0,150}
\definecolor{PrologVar}      {RGB}{105,0,175}
\definecolor{PrologComment}  {RGB}{169,082,044}
\definecolor{PrologOther}    {rgb}{0.2,0.2,0.2}
\definecolor{PrologString}   {RGB}{070,120,200}
\newcommand{\code}{\lstinline[style=MyInline]}
\lstdefinestyle{MyInline}
{
  columns=fixed,
  basicstyle = \ttfamily\color{PrologPredicate},
  breaklines = true,
  mathescape = true,
  moredelim = {*[s][\color{PrologPredicate}]{(}{)}},
  breakatwhitespace=true,
  upquote = true,
  literate =
  {?-}{{?-\,}}3
  {:-}{{:-\,}}3
}
\lstdefinestyle{MySCASP}
{
  showlines=true,
  numbersep=1em,    
  xleftmargin=0.35cm, 
  numberstyle=\tiny,
  numbers=left,
  stepnumber=1,
  breaklines = false,
  mathescape = true,
  escapechar = @,
  escapeinside = {-<}{>-},
  keywords = {},
  upquote = true,
  basicstyle = \ttfamily\relsize{-0.5}\color{PrologPredicate},
  basewidth = 0.44em,
  moredelim = {*[s][\color{PrologVar}]{(}{)}},
  moredelim = {*[s][\color{PrologString}]{'}{'}},
  moredelim = {*[s][\color{PrologOther}]{:-}{.}},
  commentstyle = \mdseries\color{PrologComment},
  escapebegin=\color{PrologVar},
  morecomment=[l]\%,
  literate     =
  {|}{{|}}2
  {&(}{{(}}1
  {&)}{{)}}1
  {.=.}{{\,\#=\,}}2
  {.<.}{{\,\#<\,}}2
  {.>.}{{\,\#>\,}}2
  {.<>.}{{\,\#\textdoublebarslash\,}}2
  {.=<.}{{\,\#=<\,}}3
  {.>=.}{{\,\#>=\,}}3
}
\usepackage{array}
\newcolumntype{L}[1]{>{\raggedright\arraybackslash}p{#1}}
\newcolumntype{C}[1]{>{\centering\arraybackslash}p{#1}}
\newcolumntype{R}[1]{>{\raggedleft\arraybackslash}p{#1}}

\definecolor{forestgreen}{rgb}{0.133, 0.545, 0.133}  

\title{An Approach to the Abstract Interpretation of\\
  Goal-Directed Answer Set Programming
  \protect
  \thanks{
Partially funded by 
MCIN/AEI 10.13039/501100011033 project COSASS (PID2021-123673OB-C32);
MICIU/AEI 10.13039/501100011033 and FEDER/EU project EVASAI (PID2024-158227NB-C32);
a research gift from Nexco Corp;
MICIU/AEI/10.13039/501100011033 Grant CEX2024-001471-M;
MICIU project CEX2024-001471-M
\emph{Mar\'{\i}a de Maeztu}; 
and by the European Union GA 101154447 NEAT. 
We also thank the anonymous
reviewers for their comments and suggestions for improvement.  } }
\author{
  \href{https://orcid.org/0000-0001-6215-1080}{DANIEL JURJO-RIVAS}
  \institute{Universidad Polit\'{e}cnica de Madrid (UPM) \& \\
    IMDEA Software Institute, Madrid, Spain}
  \email{\href{mailto:daniel.jurjo@alumnos.upm.es}{daniel.jurjo@alumnos.upm.es}}
  \and
  \href{https://orcid.org/0000-0003-4148-311X}{JOAQUÍN ARIAS}
  \institute{CETINIA, Universidad Rey Juan Carlos, Madrid, Spain}
  \email{\href{mailto:joaquin.arias@urjc.es}{joaquin.arias@urjc.es}}
  \and
  \href{https://orcid.org/0000-0001-9727-0362}{GOPAL GUPTA}
  \institute{University of Texas at Dallas, Richardson, USA}
  \email{\href{mailto:gupta@utdallas.edu}{gupta@utdallas.edu}}
  \and
  \href{https://orcid.org/0000-0001-9782-8135}{JOSE F. MORALES}
  \institute{Universidad Polit\'{e}cnica de Madrid (UPM) \& \\
             IMDEA Software Institute, Madrid, Spain}
  \email{\href{mailto:josefrancisco.morales@upm.es}{josefrancisco.morales@upm.es}}
  \and
  \href{https://orcid.org/0000-0002-1092-2071}{PEDRO LÓPEZ-GARCÍA}
  \institute{Spanish Council for Scientific Research (CSIC) \& \\
    IMDEA Software Institute, Madrid, Spain}
  \email{\href{mailto:pedro.lopez@csic.es}{pedro.lopez@csic.es}}
  \and
  \href{https://orcid.org/0000-0002-7583-323X}{MANUEL V. HERMENEGILDO}
  \institute{Universidad Polit\'{e}cnica de Madrid (UPM) \& \\
    IMDEA Software Institute, Madrid, Spain}
  \email{\href{mailto:manuel.hermenegildo@upm.es}{manuel.hermenegildo@upm.es}}
  }

\def\titlerunning{Abstract Interpretation of Goal-Directed Answer Set
Programming}
\def\authorrunning{D. Jurjo-Rivas et al.}

\begin{document}

\maketitle              
\begin{abstract}
  Abstract Interpretation infers and verifies program properties by
  over-approximating program semantics. 
  It has been highly successful for (Constraint) Logic Programming, enabling
  the analysis of determinism, types, aliasing, and resource usage, 
  as well as
  application in verification
  and program optimization. 
  However, Abstract Interpretation has not yet been studied in the
  context of Goal Directed Answer Set Programming (ASP).
  In this work, we take a first step in this direction. 
  We present a top-down algorithm based on the 
  \plai fixpoint, implemented in the abstract
  interpreter of the Ciao Prolog Preprocessor,
  to perform abstract interpretation of goal-directed \asp. 
  We also introduce the Shared-Constraints abstract
  domain, designed to capture potential relations among variables
  induced by constraints.
  Finally, we study the practicality of the approach in \scasp
  through three applications: detection of false odd loops over
  negation, efficient \textit{forall} evaluation enabled by the
  Shared-Constraints domain, and abstract specialization (including
  the simplification of required global constraints). Our results show
  that compile-time static analysis can improve the evaluation of
  goal-directed ASP programs.
\end{abstract}

\section{Introduction}
Abstract Interpretation~\cite{Cousot77} allows constructing sound
static analysis tools that can extract properties of a program by
safely approximating its semantics.  Abstract interpretation-based
analysis has been shown to be
practical and effective in the context of
(Constraint) Logic Programming
(\clp) in both verification and program optimization. 
Classic abstract interpretation is based on 
fixpoint algorithms that infer semantic information by interpreting
programs over abstract domains, 
such that the computed fixpoint 
represents a sound over-approximation of all possible concrete
executions.
Answer Set Programming (\asp), based on the stable model semantics of~\cite{gelfond88:stable_models}, has
attracted much attention due to its expressiveness, and its ability to
incorporate non-monotonicity, represent knowledge, and model
combinatorial problems.
The \scasp system implements \asp with
constraints using a goal-directed, top-down execution model that
retains logical variables both during execution and in the answer
sets.
However, Abstract Interpretation has not yet been studied in the
context of
Answer Set Programming
(see Section \ref{sec:related-work} for a discussion of related work).
Since the 
goal-directed execution model of \asp is
closer to that of \clp, it seems worthwhile to
explore whether
top-down analysis techniques via Abstract Interpretation,
such as those developed for \clp,
can be applied to \asp.
In this paper, we present an abstract interpretation framework for
goal-directed \asp which adapts the top-down abstract interpretation
framework of \plai for \clp.
We also introduce the Shared-Constraints abstract 
domain, designed to capture potential relationships among variables
induced by constraints.
%
%
We illustrate the integration of the proposed framework with \scasp
and evaluate how the information extracted at compile-time
%
can be leveraged to improve 
execution performance. Finally, we
discuss some related work and present our conclusions.

\section{Preliminaries and Notation}
\label{sec:prel}

\begin{figure}[tb]
  \begin{multicols}{4}
    \begin{lstlisting}
q(X,Y) :-
     X #> 5,
     Y = a.
r(X) :-
     X #< 1.
p(X) :-
     not q(Y,X),
     r(Y).
% DUAL
not q(X,Y) :-
     not q_1(X,Y).
not q_1(X,Y) :-
     Y \= a.
not q_1(X,Y) :-
     Y = a,
     X #=< 5.

not r(X) :-
     not r_1(X).
not r_1(X) :-
     X #>= 1.

not p(X) :-
     not p_1(X).
not p_1(X) :-
  forall(Y,not p_1(X,Y)).
not p_1(X,Y) :-
     q(Y,X).
not p_1(X,Y) :-
     not q(Y,X),
     not r(Y).
   \end{lstlisting}
  \end{multicols}
  \caption{Simple complete 
    program, including its dual rules.}
\label{code:scasp}
\vspace*{-1em}
\end{figure}
\noindent
Negation is encoded as default negation \code|not p| or as classical
negation \code{-p}.
While \code{not p} succeeds if the program cannot prove that \code{p}
holds, \code{-p} succeeds if there is a rule that explicitly states
how to deduce \code{-p}.
A \emph{constraint} is a conjunction of expressions built from
predefined predicates whose arguments are constructed using predefined
functions and variables, e.g., \code{X-Y #> 5}.
An ASP program extended with constraints under a
goal-directed execution is a set of clauses of the
form:

{\centering \code|a:- c$_a$,b$_1$,$\dots$,b$_m$,not b$_{m+1}$,$\dots$,not b$_n$.| \par}

\noindent where \code{c$_a$} is a conjunction of constraints, and
\code{a}, \code{b$_1$, $\dots$, b$_n$} are atoms.

\smallskip
\begin{definition}[Dual and Dual Program]
  The \emph{dual} of a predicate \code{p/n} is another predicate, say,
  \code{not p/n}, that returns \code{$\seq{X}$} such that
  \code{p($\seq{X}$)} is not true, i.e., it provides a constructive
  definition of \code{not p/n}.
The \emph{dual of a logic program} \prog is another logic program
containing the dual of each predicate in \prog. We say a program is
\emph{complete} if for every \code{p} $\in \prog$, \code{not p} $\in
\prog$.
\end{definition}

\vspace*{-1mm}
\noindent
To synthesize the dual of a logic program \prog, Clark’s
completion~\cite{Clark78} is computed, and then De Morgan’s laws are
applied. A detailed discussion can be found
in~\cite{scasp-iclp2018-bib,DBLP:journals/tplp/AriasEC22}.
The evaluation of dual rules requires a disequality constraint, \code{\=},
to handle the dual of
equality constraints (e.g., unifications), and a predicate \code{forall/2},
to handle the dual of existential quantifiers. Fig.~\ref{code:scasp}
shows a complete \asp program. %
Without loss of generality, we assume
programs to be complete and
predicates to be normalized.
The goal-directed, top-down evaluation of an \asp program has
similarities with SLD resolution. The computation is started by a
\emph{query}. A query is a literal
\code{?- c$_q$,l$_1$,$\dots$,l$_m$},
where the \code{l$_i$} are (possibly negated) atoms %
and \code{c$_q$} is a conjunction of constraints.
The goal-directed, top-down
evaluation %
also 
takes into account specific
characteristics of ASP and the dual programs, such as the diﬀerent
kinds of loops that may appear.

\begin{definition}[Loops and Loops Over Negation]
  A \emph{loop} is an execution trace of the form:
  $p(\seq{X}) \leadsto^{*} p(\seq{Y})$ such that
  $\solve(p(\seq{X}) = p(\seq{Y}))$ succeeds, i.e., $p(\seq{X})$
  entails/is entailed by $p(\seq{Y})$.
  If no negations are interleaved, it is called a \emph{positive
  loop}, and execution backtracks to avoid non-termination.
  Otherwise, loops are classified according to the number of
  interleaving negations:
  \begin{itemize}[nosep]
  \item \emph{Odd loop over negation}: with an odd number of
    interleaving negations. In this case execution backtracks to
    avoid contradictions of the form $p \land \lnot p$.
  \item \emph{Even loop over negation}: when there is an even number
    of negations, as in \mbox{\code{p :- not q}} \quad %
    \code{q :- not p}, multiple models are generated, such as
    \code|{p}| and \code|{q}|.
\end{itemize}
\end{definition}
\vspace*{-2mm}
\paragraph{\bfseries Abstract Interpretation.}
The main idea behind Abstract
Interpretation~\cite{Cousot77} is to interpret the program
over an abstract domain whose elements are finite
representations of possibly infinite sets of actual
states in the concrete program.
We denote the concrete domain as $D_{\gamma}$,
the abstract domain as $D_{\alpha}$, and the functions that relate
sets of concrete
states with abstract
states as the
\emph{abstraction} function $\alpha : D_{\gamma} \rightarrow D_{\alpha}$ and the \emph{concretization} function
$\gamma :D_{\alpha} \rightarrow D_{\gamma}$.
The concrete domain is typically a complete lattice with the set
inclusion order which induces an ordering relation in the abstract
domain represented by $\sqsubseteq$. Under this relation the abstract
domain is usually a complete lattice and $(D_{\gamma}, \alpha, D_{\alpha}, \gamma)$ is a Galois insertion/connection.
\emph{Top-down} analyses are a family of static analyses that build an
\emph{analysis graph} starting from a series of program \emph{entry
points}.
This graph  
is a finite abstract object whose concretization approximates the
(possibly infinite) set of %
possible executions 
of the concrete semantics. 
\vspace*{-2mm}
\section{Operational Semantics for Goal Directed ASP}
\label{sec:gandtr-char-scasp}
In this work we focus on the execution model of Goal-Directed
ASP~\cite{marple2017ALP,scasp-iclp2018-bib}. 
We represent the semantics of this model by a tree representation
similar to that used for SLD-based resolution~\cite{Lloyd87,Apt90,jaff87}.
In this representation, the execution of a program is captured by a
tree (an \andtree) which represents \emph{resolvents}.
A \emph{resolvent} is represented by $(G_1,\ldots,G_n)\cstate_i$,
where $G_i$ are literals and $\cstate_i$ is the accumulated
state resulting from the composition of the states applied so far.
A state is a tuple $\tuple{\callpath,\theta}$, where $\theta$
represents the \emph{current concrete constraint store} and \callpath
is a sequence of all calls encountered during execution (the
\emph{call path}). This path is checked prior to evaluating
predicates to avoid inconsistencies and infinite loops.%
\footnote{This is one of the key differences w.r.t. SLD resolution.}
The empty state is defined as $\emptystate=\tuple{\varepsilon,\emptyset}$
and is the initial resolvent. 
We assume a complete solver exists, and use \solve~to represent the
combination of stores.
Finally, given two states
$\localcstate_1 = \tuple{\callpath_1,\theta_1}$
and
$\localcstate_2 = \tuple{\callpath_2,\theta_2}$
their composition is defined as
  \mbox{$\localcstate_{1}\localcstate_{2} = \tuple{\callpath_1\cdot\callpath_2, \solve(\theta_1,\theta_2)}$}  
where $\callpath_1 \cdot \callpath_2$
denotes path concatenation.
To perform the computation, the leftmost literal $G_1$ is selected and
the~\emph{immediate successor} of $(G_1,\ldots,G_n)\cstate_i$ is
computed depending on $G_1 \cstate_i$ (the \emph{current goal},
denoted $G$ for short). If $G$ is a constraint,
the solver computes the \emph{immediate successor}
$(G_2,\ldots,G_n)\cstate_{i+1}$, with $\cstate_{i+1} =
\tuple{\localcstate_i,\theta_{i+1}}$ where $\theta_{i+1}$ is the
constraint store $\theta$ after solving the constraints induced
by $G$.
If $G$ is a universal quantification \code{forall(V,Goal)},
it is evaluated (for instance, by applying 
the forall algorithm~\cite{scasp-iclp2018-bib}).
On success, the \emph{immediate successor} is
$(G_2,\ldots,G_n)\cstate_{i+1}$, with $\cstate_{i+1}=\cstate_i$.%
\footnote{Note that \code{forall/2} does not update the
concrete state upon success.}

Finally, if $G$ is 
a %
predicate, the immediate successor is computed as
follows:
{\small
\begin{alignat*}{2}
  & \emptyset  && \text{If } \isloop(G,\callpath_i) \text{ returns } \odd \text{ or } \positive \\[.2em]
  & (G_2,\ldots,G_n)\cstate_{i+1}  && \text{If } \isloop(G,\callpath_i) \text{ returns } \even \\[-.4em]
  & \qquad  && \ \ \text{where } \cstate_{i+1} = \cstatetuple{\callpath_i \cdot G}{\csub_{i}}{\cstore_i} \\[.2em]
  & (B_1,\ldots,B_m,,G_2, \ldots,G_n)\cstate_{i+1}\ \ \ \ \   && \text{Otherwise } \\[-.4em]
  & \qquad  && \ \ \text{where } H \leftarrow B_1, \ldots, B_n \text{ is a clause, } \\[-.4em]
  & \qquad  && \ \ H = G \text{ succeeds with } \csub_{i+1}, \text{ and } \\[-.4em]
  & \qquad  && \ \ \cstate_{i+1} = \cstatetuple{\callpath_i \cdot G}{\csub_{i}\csub_{i+1}}{\C{i}}
\end{alignat*}
}

\noindent
where the \isloop~function inspects the call path ($\Pt{}$) in order to
detect and classify loops.
 Concrete states are represented as \andtrees, with literals in the
 state as leaves. A \emph{sequence} $\ldots,t_i,t_j\ldots$ of
 \andtrees is such that $t_j$ is the immediate successor of $t_i$.
 The set of all \andtrees which can originate from a set of queries
 completely specifies the procedural behavior of a program for that
 set of queries.

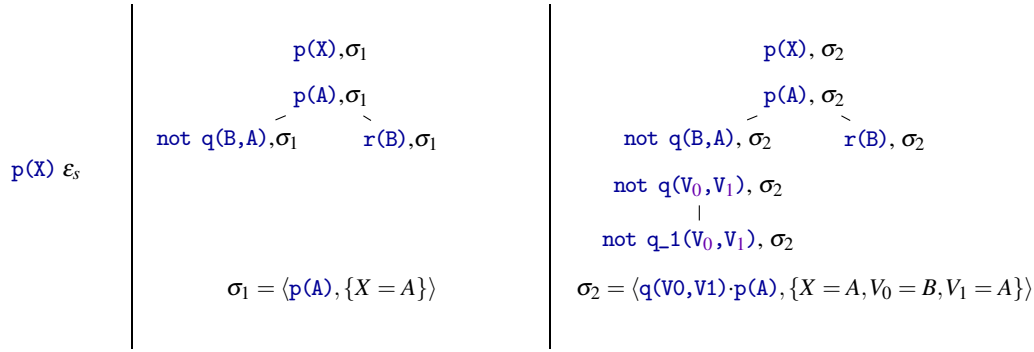
\begin{figure}[tb]
   \footnotesize
   \begin{minipage}{0.1\linewidth}
     \begin{tikzpicture}
       \node (1)  {\footnotesize\code|p(X)| $\emptystate$};
     \end{tikzpicture}
   \end{minipage}
   \vline depth 16ex height 16ex
   \hspace*{.5em}
   \begin{minipage}{0.325\linewidth}
     \begin{tikzpicture}
       \node (1)  {\footnotesize \code|p(X)|,$\sigma_1$} ;
       \node (2) [below=1mm of 1] {\footnotesize\code|p(A)|,$\sigma_1$} ;
       \node (aux) [below = 2mm of 2] {};
       \node (3) [left = 2mm of aux] {\footnotesize\code|not q(B,A)|,$\sigma_1$} ;
       \node (4) [right = 2mm of aux] {\footnotesize\code|r(B)|,$\sigma_1$} ;
       \node (theta) [below = 16mm of aux] {\footnotesize $\sigma_1 = \langle$\code{p(A)}$,\{X=A\}\rangle$};
       
       \draw[-] (2) edge (3);
       \draw[-] (2) edge (4);
     \end{tikzpicture}
   \end{minipage}
   \vline depth 16ex height 16ex
   \hspace*{.5em}
   \begin{minipage}{0.45\linewidth}
     \begin{tikzpicture}
    \node (1)  {\footnotesize\code|p(X)|, $\sigma_2$}; %
    \node (2) [below=1mm of 1] {\footnotesize\code|p(A)|, $\sigma_2$}; %
    \node (aux) [below=2mm of 2] {};
    \node (3) [left = 2mm of aux] {\footnotesize\code|not q(B,A)|, $\sigma_2$}; %
    \node (4) [right = 3mm of aux] {\footnotesize\code|r(B)|, $\sigma_2$}; %
    \node (5) [below = 1mm of 3] {\footnotesize\code|not q(V$_0$,V$_1$)|, $\sigma_2$}; %
    \node (6) [below=2mm of 5] {\footnotesize\code|not q_1(V$_0$,V$_1$)|, $\sigma_2$}; %
    \node (theta) [below = 16mm of aux] {\footnotesize$\sigma_2 = \langle$\code{q(V0,V1)}$\cdot$\code{p(A)}$,\{X=A,V_0=B,V_1=A\}\rangle$};
    \draw[-] (2) edge (3);
    \draw[-] (2) edge (4);
    \draw[-] (5) edge (6);
\end{tikzpicture}
\end{minipage}
\caption{Three consecutive \andtrees for the program in Fig.~\ref{code:scasp}}
   \label{fig:and-tree}
\end{figure}

 Fig.~\ref{fig:and-tree} shows some consecutive \andtrees for the
 program in Fig.~\ref{code:scasp} with query \code{?- p(X)}. Notice
 that all new variables are propagated during computation.

\subsection{A top-down algorithm to construct \gandtrees}
\label{sec:top-down-algorithm}

An \andtree contains more information than is necessary
for analysis, potentially involving an unbounded number of
variables. To address this in the context of \clp, Bruynooghe
\cite{bruy91} proposes the notion of \emph{\gandtrees} to capture the
computation state at each step relative to the current literal. Thus,
each node is represented as a triplet
\tuple{\pdesc{}, \ccall{}, \csucc{}},
where $\pdesc{}$ is a predicate, and
\ccall{} and \csucc{}
the call and success states over the predicate's variables.
\Gandtrees are constructed in a top-down manner starting from an initial query
$Q$, annotated with an initial call state $\ccall{0}$, whose
domain is a subset of $\vars{Q}$.
The rest of the tree is built by
expanding the leaf nodes in the following way:\\
\smallskip
\noindent
a) If \pdesc{} is a leaf adorned on the left with the call state $\ccall{i}$,
  then:
  \begin{itemize}[nosep]
  \item[i)] If $\pdesc{k}\leftarrow \qdesc{1},\ldots,\qdesc{n}$ is a
    properly renamed clause, defining \pdesc{}, then the tree is
    expanded by
    performing $\solve(\pdesc{} = \pdesc{k})$
    and adding the
    calls $\qdesc{1},\ldots,\qdesc{n}$ as children of
    \pdesc{}. \qdesc{1} is adorned on the left with the call state
    \ccall{i+1}, whose domain 
    is $\vars{\pdesc{}\leftarrow \qdesc{1},\ldots,\qdesc{n}}$.
    This tree extension is named procedure entry.
  \item[ii)] If \pdesc{} is a built-in, then the tree is expanded by
    adorning \pdesc{} on the right with a state \csucc{i},
    which is the success state of \pdesc{}.
    The domain of \csucc{i} is
    $\vars{\text{clause of } \pdesc{}}$. 
    If \pdesc{} is the last call of its clause, \csucc{i} is
    also the success state of the body; otherwise it is the call state
    of the next call.
  \end{itemize}
\noindent
b) If a node \pdesc{} is adorned on the left with a call state
\ccall{j} but not adorned with a state on the right, such that $\pdesc{}$ is
the parent of a clause body with success state $\csucc{i}$, then the
tree is expanded by adorning $\pdesc{}$ on the right with a success state
$\csucc{j}$. The domain of $\csucc{j}$ is $\vars{\text{clause of } \pdesc{}}$.
With $\pdesc{}$ the last call of its clause (the query), $\pdesc{j}$ is also the
success state of the body (the query); otherwise it is the call state
of the next call. The operation extending the tree is named \emph{procedure exit}.
\begin{definition}[Operational Semantics $\csemantics$]
\label{def:concrete-semantics}
The \emph{operational semantics} $\csemantics$ of a program $P$ for a
set of queries $Q$ is the set of all \gandtrees obtained from the
execution of $Q$ in $P$.
\end{definition}
\noindent
Given a sequence $t_1,\ldots,t_n$ of \andtrees, a corresponding
\gandtree can be built. The initial tree $t_1$, the query, corresponds
to the root node. At each step, the state captured by $t_j$
(restricted to the variables of the current literal) defines the state
of the node added to the \gandtree.
Conversely, given a \gandtree, a sequence of \andtrees can be
reconstructed. Starting from the root node, the initial tree $t_1$ is
obtained, and each subsequent tree is generated by mimicking the steps
of the \gandtree without projection, entry, or exit procedures.
 
\begin{figure}[tb]
  \renewcommand{\code}{\lstinline[style=MyInline, basewidth=0.44em]}
\small
  \begin{minipage}{0.4\linewidth}
\hspace*{-2em}    \begin{tikzpicture}
    \node (1)  {\code|p(X)|} ; \node (1a)[left =.25mm of 1] {\relsize{-2} $[\ccall{0},\csucc{0}]$} ;
    \node (2) [below=1mm of 1] {\code|p(A)|} ;
    \node (aux) [below of = 2] {};
    \node (3) [left = 7mm of aux] {\code|not q(B,A)|} ; \node (3a)[left =.25mm of 3] {\relsize{-2} $[\ccall{1}]$} ;
    \node (4) [right = 7mm of aux] {\code|r(B)|} ; \node (4a) [left =.25mm of 4] {\relsize{-2} $[\csucc{1}=\ccall{4}, \csucc{4}]$} ;
    \node (5) [below = 1mm of 3] {\code|not q(V$_0$,V$_1$)|};
    \node (6) [below of = 5] {\code|not q_1(V$_0$,V$_1$)|}; \node (5a) [left =.25mm of 5] {\relsize{-2} $[\ccall{2}, \csucc{2}]$} ;
    \node (7) [below = 1mm of 6] {\code|not q_1(V$_2$,V$_3$)|};
    \node (8) [below of = 7] {\code|V$_3$\=a|}; \node (8a) [left =.25mm of 8] {\relsize{-2} $[\ccall{3}, \csucc{4}]$} ;
    \node(9)  [below = 1mm of 4] {\code|r(V$_4$)|}; 
    \node(10)  [below of = 9] {\code|V$_4$<1|}; \node (10a) [left =.25mm of 10] {\relsize{-2} $[\Tau_7, \csucc{5}]$} ;
    \draw[-] (2) edge (3);
    \draw[-] (2) edge (4);
    \draw[-] (5) edge (6);
    \draw[-] (7) edge (8);
    \draw[-] (9) edge (10);
    \end{tikzpicture}
  \end{minipage}
  \begin{minipage}{0.55\linewidth}
    \hspace*{-2em}    \begin{tikzpicture}
      \node [ align=left, anchor=west] (t0) at (0,0.5) {\footnotesize $\ccall{0} =\emptystate$};
      \node [ align=left, anchor=west] (t1) at (0.5,0) {\footnotesize $\ccall{1} = \langle$ \code|p(A)|, $\bar{\epsilon}$, $\emptyset$ $\rangle$};
      \node [ align=left, anchor=west] (t2) at (1,-0.5) {\footnotesize $\ccall{2} = \langle$ \code|p(V$_1$)$\,\cdot\,$not q(V$_1$)|, $\bar{\epsilon}$, $\emptyset$ $\rangle$};
      \node [ align=left, anchor=west] (t3) at (1.5,-1) {\footnotesize $\ccall{3} = \langle$ \code|p(V$_3$)$\,\cdot\,$not q(V$_3$)$\,\cdot\,$not q_1(V$_2$,V$_3$)|, $\bar{\epsilon}$, $\emptyset$ $\rangle$};
      \node [ align=left, anchor=west] (t4) at (1.5,-1.5) {\footnotesize $\csucc{4} = \langle$ \code|p(V$_3$)$\,\cdot\,$not q(V$_3$)$\,\cdot\,$not q_1(V$_2$,V$_3$)|, $\bar{\epsilon}$, \code|{V$_3$\=a}| $\rangle$};
      \node [ align=left, anchor=west] (t5) at (1,-2) {\footnotesize $\csucc{2} = \langle$ \code|p(V$_1$)$\,\cdot\,$not q(V$_1$)$\,\cdot\,$not q_1(V$_0$,V$_1$)|, $\bar{\epsilon}$, \code|{V$_3$\=a}| $\rangle$};
      \node [ align=left, anchor=west] (t6) at (0.5,-2.5) {\footnotesize $\csucc{1}=\ccall{4} = \langle$ \code|p(A)$\,\cdot\,$not q(A)$\,\cdot\,$not q_1(B,A)|, $\bar{\epsilon}$, \code|{A\=a}| $\rangle$}; 
      \node [ align=left, anchor=west] (t7) at (1,-3) {\footnotesize $\Tau_7 = \langle$ \code|p(_)$\,\cdot\,$not (V$_4$,_)$\,\cdot\,$not q_1(V$_4$,_)|, $\bar{\epsilon}$, $\emptyset$ $\rangle$};  
      \node [ align=left, anchor=west] (t8) at (1,-3.5) {\footnotesize $\csucc{5} = \langle$ \code|p(_)$\,\cdot\,$not q(V$_4$,_)$\,\cdot\,$not q_1(V$_4$,_)|, $\bar{\epsilon}$, \code|{V$_4$<1}| $\rangle$};
      \node [ align=left, anchor=west] (t9) at (0.5,-4) {\footnotesize $\csucc{4} = \langle$ \code|p(A)$\,\cdot\,$not q(B,A)$\,\cdot\,$not q_1(V$_4$,A) r(B)|, $\bar{\epsilon}$, \code|{A\=a, B<1}| $\rangle$};
      \node [ align=left, anchor=west] (t10) at (0,-4.5) {\footnotesize $\csucc{0} = \langle$ \code|p(X)$\,\cdot\,$not (_,X)$\,\cdot\,$not q_1(_,X) r(_)|, $\bar{\epsilon}$, \code|{A\=a}| $\rangle$};

      \draw[->,blue] (t0.south west) -- ++(0,0) |- (t1.west);
      \draw[->,blue] (t1.south west) -- ++(0,0) |- (t2.west);
      \draw[->,blue] (t2.south west) -- ++(0,0) |- (t3.west);
      \draw[->,blue] (t6.south west) -- ++(0,0) |- (t7.west);
      \draw[->,forestgreen] (t4.west) -- ++(0,0) -| (t5.north west);
      \draw[->,forestgreen] (t5.west) -- ++(0,0) -| (t6.north west);
      \draw[->,forestgreen] (t8.west) -- ++(0,0) -| (t9.north west);
      \draw[->,forestgreen] (t9.west) -- ++(0,0) -| (t10.north west);
    \end{tikzpicture}
  \end{minipage}
  \caption{Complete \gandtree for the program in Fig.~\ref{code:scasp} and its construction steps.}
  \label{fig:gand-tree}
  \centerline{\relsize{-1} NOTE: Blue arrows represent procedure entries while green
    arrows represent procedure exits.}
  \vspace*{-.5em}
\end{figure}
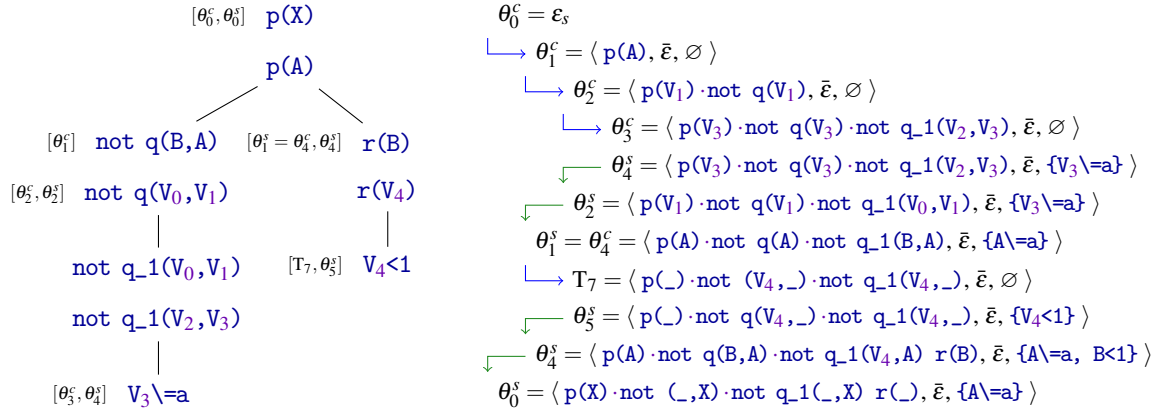

\begin{example}[\gandtree]
Fig.~\ref{fig:gand-tree} presents a complete \gandtree for the
program in Fig.~\ref{code:scasp} and the query \code{?- p(X)}.
Starting with call store $\ccall{0} =\emptystate$ the execution succeeds
with store
\mbox{$\csucc{0} =$ \code|$\langle\;$p(X)$\,\cdot\,$not q(_,X)$\,\cdot\,$not q_1(_,X)$\,\cdot\,$r(_)|}, $\bar{\epsilon}$, \code|{X\=a}$\;\rangle$|.  
From the tree, it is possible to recover \code+{$~$p(X|{X\=a}),+ \code+not q(Y, X|{Y#<1,X\=a}), r(Y|{Y#<1})$~$}+, the only answer set generated by this program for the given query.
Notice how $\ccall{0},\ccall{1}$ and $\ccall{2}$ correspond to projecting $\emptystate, \sigma_1,\sigma_2$ in the sequence of \andtrees in~Fig.~\ref{fig:and-tree}.
\end{example}
\section{An Abstract Interpreter for Goal-Directed ASP%
}
\label{sec:an-abstr-interpr}
In Section~\ref{sec:gandtr-char-scasp} we presented the operational
semantics for goal directed \asp. In this section we present a
top-down approach for the analysis of goal-directed \asp programs.
The objective of this analysis is to build an \emph{analysis graph}
starting from a series of program entry points. 

\smallskip
\begin{definition}[Analysis Graph]
  An analysis graph is a (directed) call graph and a mapping function
  from predicate descriptors and call states to success states.
  A \emph{node} is a tuple $\langle \pdesc{},\lambda^c \rangle$,
  representing that a call to a predicate $\langle \pdesc{},\lambda^c
  \rangle$ is possibly made in the concrete program execution and has
  an associated success abstraction $\lambda^s$ through the mapping
  $\langle \pdesc{},\lambda^c \rangle \mapsto \lambda^s$.
  An \emph{edge} in an analysis graph is of the form
$\langle A,\lambda^c \rangle \rightarrow_{k,i} \langle B,\eta^c\rangle$. This
represents that calling predicate $A$ with calling pattern $\lambda^c$
may cause $B$ to be called (via the i-th literal in the k-th clause of
$A$, $A_{i,k}$) with calling pattern $\eta^c$.
\end{definition}

\smallskip\noindent
The analysis graph is built to capture the behaviour of the original
program by abstracting the set of \gandtrees of $P$ with a given set
of queries $\Q$. 

\subsection{A PLAI-based abstract interpreter for goal-directed ASP}
\label{sec:plai-based-abstract}

In order to construct the analysis graph, we modify the \plai~\cite{mcctr-fixpt,ai-jlp, anconsall-acm} top-down
analysis framework, to adapt to the semantics of goal-directed ASP.
The algorithm (which is parametric to a given abstract domain
$D_{\alpha}$)
receives as input a program $P$ and a set of initial
\emph{abstract queries},
$Q_{\alpha}={\langle A_i,\lambda^c_{i}  \rangle}$,
where each $A_i$ is an atom and $\lambda^c_{i}\in D_{\alpha}$.
Algorithm~\ref{alg:analyze-scasp} presents the analysis procedure.
\begin{figure}
  \algrenewcommand\alglinenumber[1]{\tiny #1:}
  \algrenewcommand\algorithmicindent{0.9em} 
  \begin{minipage}[t]{.44\textwidth}
  \begin{algorithmic}[1]
    \begin{scriptsize}
      \Function{Analyze}{\prog,\aqueries}
    \State \G $\gets \emptyset$
    \For{\tuple{\pdesc{},\acall{}} $\in$ \aqueries} \label{alg:queries-beg}
        \State \G $\gets$ \Call{CallToSucc}{\pdesc{},\acall{},\G,$\emptyset$}   
    \EndFor \label{alg:queries-end}
    \State \Return \G
    \EndFunction
    
    \Function{CallToSucc}{\pdesc{}, \acall{},\G,\trace} \label{alg:c2s-beg}
    \State \proj $\gets$ \aproject(\vars{\pdesc{}}, \acall{})  \label{alg:ana-proj}
    \State \texttt{Type} $\gets$ \Call{TypeOfLoop}{\pdesc{},\trace} \label{alg:ana-typeofloop}
    \If{(\texttt{Type} = \even)}\label{alg:ana-even-loop-beg}
         \State \head $\gets$ \Call{FirstApp}{\pdesc{},\proj,\trace}
         \State \entry $\gets \mbox{\Call{\calltoentry}{\proj, \pdesc{}, \head}}$
         \State \aprime{} $\gets \mbox{ \Call{\exittoprime}{\entry, \head, \pdesc{}}}$
         \State \succ $\gets$ \Call{\extend}{\acall{},\aprime{}}\label{alg:ana-even-loop-end}
      \ElsIf{(\texttt{Type} $\in \{\noloop,\texttt{PossOdd}\}$)}\label{alg:ana-no-loop-beg}
          \State \trace $\gets$ \Call{append}{\pdesc{},\proj,\trace} \label{alg:c2s-append}
          \State \mbox{\aprime{} $\gets$ \Call{AnalyzePred}{\proj,\tuple{\pdesc{},\acall{}},\G,\trace}}\label{alg:analyze-pred-call}
          \State \succ $\gets$ \Call{\extend}{\acall{},\aprime{}}
          \State \texttt{Trace} $\gets$ \Call{removehead}{\texttt{Trace}}
      \ElsIf{(\texttt{Type} $\in\{\odd,\positive\}$)} \label{alg:ana-odd-loop-beg}
      \State \succ $\gets \bot$\label{alg:ana-odd-loop-end}
      \ElsIf{(\texttt{Type} = \texttt{PossEven})}\label{alg:ana-may-even-loop-beg}
         \State \head $\gets$ \Call{FirstApp}{\pdesc{},\proj,\trace}
         \State \entry $\gets \mbox{\Call{\calltoentry}{\proj, \pdesc{}, \head}}$
         \State \mbox{$\aprime{1} \gets \Call{\exittoprime}{\entry, \head, \pdesc{}}$}
         \State \mbox{$\aprime{2}\gets$ \Call{AnalyzePred}{\proj,\tuple{\pdesc{},\acall{}},\G,\trace}}
         \State \mbox{\asucc{}$\gets$ \Call{\extend}{\acall{},$\aprime{1}\sqcup\aprime{2}$}\label{alg:ana-may-even-loop-end}}
      \EndIf
      \vspace*{-.2em}
      \State \G $\gets$ \update{\G}{\map{\pdesc{}}{\acall{}}{\asucc{}}} \label{alg:update-succ}
      \State \Return \asucc{},\G
      \EndFunction \label{alg:c2s-end}
      \end{scriptsize}
\end{algorithmic}
\end{minipage}
\begin{minipage}[t]{.55\textwidth}

  \begin{algorithmic}[1]
    \setcounter{ALG@line}{28} 
    \begin{scriptsize}
      \Function{AnalyzePred}{\proj,\tuple{\pdesc{},\acall{}},\G,\trace}\label{alg:analyze-pred-beg}
      \vspace*{-.4em}
     \If{\fmap{\pdesc{}}{\proj}{\aprime{}} $\in$ \G}
         \State \Return \aprime{}
     \Else
         \State $\aprime{} \gets \bot$
         \Repeat \label{alg:repeat-beg}
         \State $\aprime{0} \gets \aprime{}$
            \For{$\pred{\pdesc{k}}{\{\pdesc{k,j}\}} \in \prog$}                                                   
               \State \mbox{\entry $\gets$ \Call{\calltoentry}{\proj, \pdesc{}, \pdesc{k}}}\label{alg:call-to-entry}
               \State \mbox{\entry $\gets$ \Call{\augment}{\vars{\pdesc{k,j}}$\backslash\vars{\pdesc{k}}$, \entry}}\label{alg:augment}
               \State \mbox{\exit, \G$\gets$ \Call{\entrytoexit}{\entry,$\pdesc{k},\{\pdesc{k,j}\}$,\tuple{\pdesc{},\acall{}},\trace,\G}}\label{alg:entry2exit-call}
               \State \mbox{$\aprime{1} \gets$ \Call{\exittoprime}{\project(\vars{$\pdesc{k}$}, \exit), \pdesc{k}, \pdesc{}}}\label{alg:exit-to-prime}
               \State \aprime{} $\gets\aprime{}\sqcup\aprime{1}$          
             \EndFor
      \vspace*{-.4em}
         \State \G $\gets$ \update{\G}{\fmap{\pdesc{}}{\proj}{\aprime{}}}
         \label{alg:update-fixp}
         \Until{$\aprime{0} = \aprime{}$} \label{alg:repeat-end}
      \vspace*{-.4em}
         \State \G $\gets$ \del{\G}{\fmap{\pdesc{}}{\proj}{\aprime{}}}
         \State \Return \aprime{}
     \EndIf    
     \EndFunction\label{alg:analyze-pred-end}

     \Function{\entrytoexit}{$\entry, \pdesc{k},\{\pdesc{k,j}\},\tuple{\pdesc{},\acall{}}, \trace, \G$}\label{alg:ana-entry2exit-beg}
     \State $\exit \gets \entry$ \label{alg:ana-entry2exit-init}
     \For{$\pdesc{k,i} \in \{\pdesc{k,j}\}$}\label{alg:ana-entry2exit-loop}
     \If{$\pdesc{k,i} \in \prog$} \label{alg:entry2exit:predicate-beg}
          \State \mbox{\G $\gets$ \update{\G}{\edge{\pdesc{}}{\acall{}}{\pdesc{k,i}}{\exit}} \label{alg:update-call}}
          \State \exit $\gets$ \Call{CallToSucc}{\pdesc{k,i}, \exit,\G,\trace} \label{alg:entry2exit:predicate-end}
         \ElsIf{$\Call{IsForall}{\pdesc{k,j}}$} \label{alg:ana-forall-beg}
             \State $\qdesc{\forall} \gets \Call{GetForallGoal}{\pdesc{k,i}}$
             \State $\acall{\forall} \gets \Call{TopMostCall}{\qdesc{\forall}}$
             \If{$\Call{CallToSucc}{\qdesc{\forall},\acall{\forall},\emptyset,\trace}=\bot$}
                \State $\exit\gets\bot$
             \Else
                \State $\exit\gets\acall{}$
             \EndIf \label{alg:ana-forall-end}
      \vspace*{-.3em}
          \Else
              \State $\proj \gets \Call{\aproject}{\vars{\pdesc{k,i}}, \exit}$ \label{alg:entry2exit:builtin-beg}
      \vspace*{-.2em}
              \State \mbox{$\exit \gets \Call{\extend}{\exit, \builtin(\pdesc{k,i}, \proj)}$ \label{alg:entry2exit:builtin-end}}
          \EndIf
      \EndFor\label{alg:ana-entry2exit-loop-end}
      \vspace*{-.2em}
      \State \Return $\exit$,\G \label{alg:ana-entry2exit-end}
\EndFunction
     \end{scriptsize}
  \end{algorithmic}
\end{minipage}
   \caption{A schematic description the analysis algorithm}\label{alg:analyze-scasp}
\end{figure}
For each abstract query, the function \textsc{CallToSucc} is invoked
to compute the success abstraction of the corresponding abstract call
pattern.
First, it determines whether the current call corresponds to a loop
via \textsc{TypeOfLoop} (Line~\ref{alg:ana-typeofloop}).
This function inspects the current trace and determines whether there
will always occur a loop (\even, \odd, \positive) or whether there may
occur a loop.
Since the abstract state over-approximates
the concrete state,
in most cases it cannot be determined whether a loop is \even or \odd.%
\footnote{In fact, this is only possible if both literals in the trace are ground.}
Therefore, loop detection is performed syntactically: if loops are
possible, the analysis has to assume the least precise (most general)
option to happen.
Once the case is decided, the analysis proceeds as follows:
\begin{itemize}[nosep]
\item If there are \odd or \positive loops, the abstract success
  ($\asucc{}$) is assigned to $\bot$. This reflects that, for any
  corresponding concrete derivation, the execution will
  fail (Lines~\ref{alg:ana-odd-loop-beg}--\ref{alg:ana-odd-loop-end}).
\item It there are \even loops the success abstraction is the result
  of abstractly
  joining the call with the looping variant
  (Lines~\ref{alg:ana-even-loop-beg}--\ref{alg:ana-even-loop-end}).
\item If there may be even loops, two prime abstractions are
  computed. One assuming the even case (by
  performing the abstract
  equalities), 
  and another assuming that there is no even loop:
  continuing with the analysis. Then, the success abstraction is the
  result of extending with the least upper bound of both abstractions
  (Lines~\ref{alg:ana-may-even-loop-beg}--\ref{alg:ana-may-even-loop-end}).
\item If there are no loops (or there may be \odd loops), analysis proceeds
  into \textsc{AnalyzePred}.
\end{itemize}
 
\noindent The function \textsc{AnalyzePred}
(Lines~\ref{alg:analyze-pred-beg}--\ref{alg:analyze-pred-end})
computes the abstract success of a given predicate. It iterates over
\emph{all} the clauses of the predicate,%
\footnote{Since given a literal we cannot know (in general) which
clause will be executed.}
computing for each a prime abstraction
(the abstract state after executing the clause body) by calling
\textsc{EntryToExit}.
The prime abstractions of all the clauses are joined via $\sqcup$ and the
result is stored in $\mathcal{G}$.
This is repeated until a fixpoint is reached
(Lines~\ref{alg:repeat-beg}--\ref{alg:repeat-end}).  During the
computation of the fixpoint, auxiliary edges (\sfmap{\cdot}{\cdot})
are temporarily added to the graph.
Function \textsc{EntryToExit} traverses the body of each clause. For
each predicate, an edge is added to $\G$ (Line~\ref{alg:update-call})
and then \Call{CallToSucc}{} is invoked recursively.
For a \code{forall} literal, the inner goal is analyzed independently
with a top-most call abstraction, since the concrete execution is
performed with new fresh variables. If the result is $\bot$, meaning
that the \code{forall} invocation fails for any concrete execution, all
the subsequent literals are unreachable and the abstraction is set to
$\bot$ (Lines~\ref{alg:ana-forall-beg}--\ref{alg:ana-forall-end}).
Finally, for built-ins and constraints, the abstract state is updated by means
of \builtin (\ref{alg:entry2exit:builtin-beg}--\ref{alg:entry2exit:builtin-end}).
\vspace*{-2mm}
\subsection{Shared-Constraints: An Abstract Domain for ASP + Constraints}
\label{sec:shar-constr-an}

Similarly to the \emph{set-sharing} domain %
\cite{abs-int-naclp89,jacobs89} for variable aliasing,
we introduce the 
\emph{Shared-Constraints} abstract domain
which 
encodes which variables
\emph{may be} constrained at runtime by capturing them within the same
\emph{sharing set}.
The lattice of the \emph{Shared-Constraints} abstract domain is defined as
$S_{ct}= \mathcal{P}(\mathcal{P}(PV)),$ where $PV$ is the set of
program variables, equipped with the order relation:
\mbox{$Ss_1 \leq Ss_2 \Leftrightarrow \forall S_1\in Ss_1,\exists S_2\in Ss_2 \text{ s.t. } S_1\subseteq S_2$}.
Thus, given the constraint store $\{X > R, R > Y, Z \neq a\}$, its
abstraction %
is $\{\{X,Y\}, \{Z\}\}$. It captures that $X$ may be constrained with
$Y$, while $Z$ is not constrained with either $X$ or $Y$.
We consider the constraint relation between variables to be
transitive. For instance, if $\{X,Y\}, \{Z,T\}$ are sharing sets, and
$Z$ is constrained with $X$, all variables become mutually
constrained, resulting in the sharing set $\{X,Y,Z,T\}$.
The abstraction of built-in 
literals
is done by means of the \builtin function defined as:
\mbox{$\builtin(\literal, \asub{}) =  \;\{ S \in \asub{} \mid S \cap \vars{\literal} = \emptyset \} \; \cup \;\{ S \in \asub{} \mid S \cap \vars{\literal} \neq \emptyset \}$.}
When encountering
such literals, we conservatively assume
that all their variables may become constrained and incorporate this
information into the abstraction. For constraints, this definition
yields a
precise abstraction.
For other literals, it provides
a safe over-approximation. Precision can be improved by specializing
the treatment of selected built-ins and operators. For example, for a
call to \texttt{length}$(L,N)$, it can be safely inferred that $L$ and
$N$ are not mutually constrained.
The definitions of the \calltoentry and \exittoprime operations differ
only in the order of
equalities. In particular, \calltoentry is
defined as
 $  \calltoentry(\asub{},T_1,T_2) = \amgu(T_{1}=T_{2}, \asub{})$
where the
function \amgu first computes
the solved form of the
equality $T_{1}=T_{2}$ and, from this set of equations,
invokes \builtin iteratively until obtaining the
resulting abstract constraint.
\noindent
The remaining operations are defined as:
$ \asub{1}\sqcup \asub{2} = \asub{1} \cup \asub{2}$, 
$ \aproject(\asub{},Vs) = \{ Ss \cap Vs \mid Ss\in\asub{} \}$, 
$ \extend(\asub{1},\asub{2}) = \{ Ss \cup Pr \mid Ss \in \asub{1} \wedge Pr=\bigcup\{P \in \asub{2} \mid P \cap Ss \neq \emptyset \} \}$, and 
$ \augment(\asub{},Vs) = \asub{}$.
The \augment~operation requires its input variables to be fresh and
unconstrained; therefore, the abstraction remains unchanged.

\smallskip
This abstract domain allows capturing dependencies between variables
by recording which variables may become mutually constrained during
goal evaluation.  The resulting information is used in
Section~\ref{sec:eval-forall} to reorganize nested \texttt{forall}
literals at compile time by grouping universally quantified variables
that belong to the same dependency class.
\section{Some Applications of the Abstract Interpretation of Goal-Directed ASP}
\label{sec:appls}
This section evaluates the techniques presented and describes
applications of abstract interpretation-based analyses in the context
of goal-directed \asp.
First, %
we propose an
Abstract
Interpretation-based transformation for efficient forall evaluations in \scasp
using the Shared-Constraints abstract domain.
Then, %
we discuss how abstract
specialization of goal-directed \asp programs can be perfomed by using
already mature abstract domains in the context of \clp.
Finally, we use the information inferred by the analyzer to detect
``false'' odd loops over negation, enabling the compiler to skip generating the
corresponding global constraints, and %
avoiding %
unnecessary
consistency checks.
Experiments %
were conducted on a
MacBook Air M2 with 16 GB of RAM and 500 GB.
\subsection{Efficient forall Evaluation enabled by the Shared-Constraints Domain}
\label{sec:eval-forall}
One of the main motivations for introducing abstract interpretation in
the context of goal-directed \asp is the possibility of extracting
compile-time information that can be exploited to improve the
execution of the programs.
One of the most computationally expensive steps in the execution of
\scasp programs is the evaluation of forall predicates.
The algorithm to handle \code{forall(V,G)}, initially proposed
in~\cite{marple2017ALP} %
and extended 
in~\cite{scasp-iclp2018-bib} to handle constraints,
proceeds
by iteratively narrowing the constraint store $C_i$ under which the
goal \code{G} is being evaluated (initially the constraint store has no
restriction on the variable \code{V}, i.e., it is free).
In each iteration, when an answer $A$ is found, the
subsequent constraint store $C_{i+1}$ is
$C_{i} \wedge A_V \wedge \neg A_{\overline{V}}$, where $A_V$ is the
projection of $A$ on \code{V} and $A_{\overline{V}}$ is the projection
of $A$ onto the rest of variables on \code{G}.
This procedure is computationally expensive, as an answer/constraint
store may contain multiple constraints whose negation/dual must be
considered. Concretely, given a store containing $n$ constraints,
there are up to $2^n - 1$ distinct combinations of dual constraints
that must be explored during the evaluation. E.g., given
\code|{A>5,A<10,B>=0}|, its $7$ duals are: \code|{A=<5,A<10,B>=0}|,
\code|{A>5,A>=10,B>=0}|, \code|{A>5,A<10,B<0}|, \code|{A=<5,A=>10,B>=0}|,
etc.
\noindent
The initial algorithm
was based on the assumption that there was only a single constraint domain
(the Herbrand domain) and that constraints between free variables
were not allowed. Under these assumptions, the \code{forall/2} predicate over multiple
variables %
was evaluated in a nested manner.\footnote{In fact, the \scasp
compiler still preserves this representation, i.e.,
\code{forall(A,forall(B, goal(A,C,B)))}.} This algorithm was easily adapted to
handle constraints over the rationals/reals, %
provided that
these constraints did not involve relations between free variables at
the time the forall was evaluated.
For this reason, the constraints had to be placed at the end of the
rule body, leaving only one variable free.
The extended algorithm for constraints (\code{c_forall}) 
was implemented by collecting all the variables in a single set,
transforming the previous nested call into a form such as
\code{forall([A,B], goal(A,C,B))}. This implementation improves upon
the initial algorithm by supporting constraints among free variables
over the rationals/reals. E.g., rules of the form \code{p(a,X,Y) :- X < Y}
are allowed, despite both \code{X} and \code{Y} being unbound
at the time the constraint is invoked during the forall evaluation.
However, to avoid the combinatorial explosion due to the dual
constraint mentioned earlier, \code{c_forall} assumes that the free
variables are restricted among them in a unique cluster, i.e., rules
of the form \code{p(X,Y,Z,T) :- X < Y, Z < T.} are not
supported. Under this assumption the dual of a constraint store can be
``optimized'' to only generate $n$ distinct dual constraints stores,
for instance \code|{A>5,A<10,B>=A}| has only three dual stores:
\code|{A=<5,A<10,B>=A}|, \code|{A>5,A>=10,B>=A}|, and
\code|{A>5,A<10,B<A}|.

\vspace*{-0.5em}
\paragraph {\bfseries Exploiting the Shared-Constraints Abstract Domain.}

The information inferred by the abstract domain Shared-Constraints,
described in Section~\ref{sec:shar-constr-an}, generates clusters of
variables that may be constrained at execution time.
This allows the compiler 
to reorganize nested \code{forall} predicates at compile time by
grouping universally quantified variables that belong to the same
dependency class.
Then, at execution time, each cluster is handled in a nested manner,
following the algorithm strategy proposed in~\cite{marple2017ALP},
while still supporting constraints among free variables within each
cluster as implemented in~\cite{scasp-iclp2018-bib}.
The compile-time procedure starts by selecting each (possibly nested)
\code{forall} predicate in the program. Then, the innermost goal is
analyzed starting from an empty abstraction. The usage of an empty
abstraction is sound because universally quantified variables are
unconstrained when the \code{forall} is invoked.
Consider the previous rule, \code{p(X,Y,Z,T) :- X < Y, Z < T.}, and the
nested forall predicate
\code{forall(X,forall(Y,forall(T,p(X,Y,Z,Y))))}.
The analysis invokes the abstract interpreter on the inner goal:  
\mbox{$\Call{Analyze}{\prog, \tuple{p(X,Y,Z,T), \emptyset}}.$}
The resulting success abstraction for \code{p(X,Y,Z,T)} groups
variables that may become mutually constrained, e.g., \code|[[X,Y], [T,Z]]|.
Based on this abstraction, nested universal quantifiers are
automatically reorganized by grouping quantified variables that belong
to the same dependency class,
obtaining 
\code{forall([X,Y], forall([T], p(X,Y,T,Z)))},
preserving the declarative semantics of the program and improving its evaluation.
\begin{table}[tb]
  \centering
  \caption{Runtime (in ms.) for Total / Answers evaluation and Abstract Interpretation.}
  \label{tab:forall_times}
  \footnotesize
    \begin{tabular}{L{12mm} p{2mm} R{9mm}R{17mm} p{3mm} R{10mm}R{17mm} p{3mm} R{10mm}R{15mm} R{10mm} }  
    \toprule
    &
    & \multicolumn{2}{c}{\code|prev_forall|} &
    & \multicolumn{2}{c}{\code|c_forall|} &
      & \multicolumn{2}{c}{\code|cls_forall|} &
\multirow{2}{*}{A.I.}
      \\
    \cmidrule(l){3-4}     \cmidrule(l){6-7}    \cmidrule(l){9-10} 
    &  & Total & Answers &  & Total   & Answers &   & Total   & Answers    \\
\midrule
light-1 &  & 280 & 48  \ (1)      & &  850   & 274     (1)       &   & 350    & 59   \ (1) &   9            \\
light-2 &  & 310 & 81  \ (1)      & & 1940   & 383     (1)       &   & 770    & 86   \ (1) &  10            \\
light-3 &  & 200 & -   \ (0)      & &  190   & -       (0)       &   & 470    & -    \ (0) &  10            \\
light-4 &  & 420 & 160 \ (1)      & & \multicolumn{2}{c}{wrong}  &   & 1060   & 196  \ (1) &  11            \\
tap-1   &  & 680 & 43  \ (1)      & & 263350  & 223841 (1)       &   & 1000   & 42   \ (1) &  18            \\
tap-2   &  & 240 & 28  \ (1)      & & 1270    & 492    (1)       &   & 590    & 26   \ (1) &  18            \\
tap-3   &  & 280 & 60  \ (2)      & & 2370    & 1007   (2)       &   & 720    & 61   \ (2) &  19            \\
\midrule                                                                                         
graph-0 &  & 15939 & 14322   (30) & & 1400    & 1162  (3)      &   & 1910   & 1210  \ (3) &   4            \\
graph-1 &  & 8950  & 8096    (30) & & 790     & 569   (3)      &   & 8950   & 8646  (30) &   8            \\
hanoi   &  & 250   & 55   \   (1) & & 240     & 53    (1)      &   & 360    & 52    \ (1) &   1            \\
\bottomrule
  \end{tabular}
  \centerline{\footnotesize NOTE: In parenthesis the number of Answers.}
\end{table}

\paragraph{\bfseries Validation.}
We validate the benefits of using the Abstract
Interpreter with the Shared-Constraints domain in the execution of the
\code{c_forall} predicate. As a baseline, we present the
execution times of the \code{prev_forall} predicate, which is the
fastest since it does not consider any kind of relationships among
variables.
We selected a suite of benchmarks%
: (i) the \code{light-$j$} and \code{tap-$k$} examples describe properties of a
time-varying light and tap flows, both
modeled in~\cite{DBLP:journals/tplp/AriasEC22} under  Event
Calculus, featuring complex \code{forall} predicates; (ii)
the remaining examples include two implementations of the
graph-coloring problem, and the Tower of Hanoi.
These last examples do not involve complex \code{forall} invocations
or constraints and are used to evaluate the overhead introduced by the
abstract interpretation approach.
Table~\ref{tab:forall_times} shows evaluation times for these
examples, comparing the three \code{forall} implementations. For each
example, we report the total execution time (Total, in ms), the time
spent computing answers (Answers), the number of answers (in
parenthesis), and, for the \code{cls_forall} the overhead due to the
static analysis and the forall rewriting (AI).
The results show that detecting variable independence under constraints
can significantly improve performance. In particular,
the \code{tap-1} example exhibits a roughly \textbf{250$\times$ speed-up}.
The program contains 12
\code{forall} predicates, 11 of which quantify over at least two
variables, while only 3 involve variables that may become mutually
constrained at run time, 
while the remaining can 
be treated independently.
In the cases of graph-0, grouping universal variables leads to a
different number of answers. All of these answers are correct,
although in some cases they are redundant.
A similar effect is observed when the \code{prev_forall} predicate is
used, which suggests that this treatment of universally quantified
variables is more inclined to generating redundant answers. Moreover, we
observe that the optimization often brings execution times close to
those obtained with \code{prev_forall}. This highlights that, while in
some cases grouping variables has little apparent effect, in others
even a small reduction in the number of \code{forall} invocations
can lead to substantial performance improvements.
%
The additional
compilation effort is negligible, remaining below 20 milliseconds in
all benchmarks.

\begin{figure}[tb]
  \begin{minipage}{0.3\linewidth}
  \begin{lstlisting}
%%% Graph Coloring %%%
node(1).       node(2).
node(3).       node(4).
node(5).
edge(1,2).   edge(1,3).
edge(2,4).   edge(2,5).
edge(3,4).   edge(3,5).
color(red).
color(green).
color(yellow).

other_color(X,C) :-
    color(C), color(C2),
    C \= C2, color(X,C2).
color(X,C) :-
    node(X), color(C),
    not other_color(X,C).

:- edge(X,Y), color(X,C), 
                    color(Y,C).
  \end{lstlisting}
  \end{minipage}
  \hfill
  \vline depth 20ex height 20ex
  \hfill
  \begin{minipage}{0.63\linewidth}
    \vspace*{-0,75em}
    \begin{multicols}{2}
      \begin{lstlisting}
%%% Specialization %%%
other_color(1,green) :-
     color(1,red).
other_color(1,green) :-
     color(1,yellow).
other_color(1,red) :-
     color(1,green).
other_color(1,red) :-
     color(1,yellow).
$\dots$
% Some rules are omitted
$\dots$
other_color(5,red) :-
     color(5,yellow).
other_color(5,yellow) :-
     color(5,green).
other_color(5,yellow) :-
     color(5,red).
color(1,green) :-
     not other_color(1,green).
color(1,red) :-
     not other_color(1,red).
color(1,yellow) :-
     not other_color(1,yellow).
color(2,green) :-
     not other_color(2,green).
$\dots$
color(4,green) :-
     not other_color(4,green).
color(4,red) :-
     not other_color(4,red).
color(4,yellow) :-
     not other_color(4,yellow).
color(5,green) :-
     not other_color(5,green).
color(5,red) :-
     not other_color(5,red).
color(5,yellow) :-
     not other_color(5,yellow).
\end{lstlisting}
   \end{multicols}
 \end{minipage}
 \vspace*{-.25em}
\caption{Implementation of the Graph Coloring Problem and its specialization using \emph{\eterms}.}
\label{code:graph-color}
\vspace*{-1em}
\end{figure}

\subsection{Toward Abstract Specialization}
\label{sec:peval}
Since the approach in
Section~\ref{sec:an-abstr-interpr} adapts the \plai\ framework to
analyze goal-directed, top-down \asp\ programs and does not require
the underlying abstract domain to introduce additional operations, it
allows for the direct use of domains designed for \clp.
However, custom domains and algorithms could yield more precise
abstractions.
We focus on the \eterms abstract domain~\cite{eterms-sas02-bib},
which implements a type abstraction with efficient widening, providing
a precise over-approximation of regular types.
This captures all possible
Herbrand equalities in any successful answer set,
enabling partial grounding tailored to specific queries.
Compared to query-independent analyses, the top-down, goal-directed
approach produces more concrete %
constraints, since it takes into
account only the possible %
constraints from an initial set of queries.
The procedure for performing specialization using this abstract
information consists of two steps: (i)~\textbf{Analysis:} the program
is analyzed using the \eterms domain. At success, each variable’s
Herbrand domain is over-approximated across all possible answer sets.
(ii)~\textbf{Specialization:} using this information, the program is
specialized, generating rule instances suitable for abstract
evaluation.

\vspace*{-1em}
\paragraph{\textbf{Validation.}}
Fig.~\ref{code:graph-color} shows, on the left, an instance of the
graph coloring problem, and on the right the resulting program after
applying abstract specialization using the information inferred by
\emph{\eterms}.
For the specialized program, 150 answers to \code{?- color(X,Y)} were
computed in 65 seconds (including 225 ms precompilation), with a mean
answer time of 433 ms. The default program required 80 seconds, with a
mean answer time of 533 ms. This represents an 18\% reduction
in execution time.
\subsection{Detection of False Odd Loops over Negation}
\label{sec:olon}
During compilation, s(CASP) generates denials for OLON rules, which
are then checked by the predicate \code{nmr_check} to prevent
inconsistencies.
For instance, the rule \code{p(X) :- q(X,Y), not p(Y).} is treated as
an OLON rule when the possible values of \code{X} and \code{Y} are
unknown. However, if the program only includes \code{q(a,b)} for
\code{q/2}, no denial is required.
We use type information from the \emph{\eterms} abstract domain to
filter candidate NMR rules in four steps: (i) the program is analyzed
without NMR checks; (ii) for each OLON rule, the types of its initial
and final literals are intersected: a non-empty intersection indicates
a potential loop, while an empty one means it cannot occur; (iii) to
ensure safety, all reachable literals
matching the initial
literal are also checked; (iv) if all intersections are empty, the
OLON loop can be safely omitted from inconsistency checks.
\begin{figure}[tb]
  \begin{minipage}{0.23\linewidth}
\begin{lstlisting}
%%% path_disc/2 %%%
initial_node(a).

edge(a,b). 
edge(a,c).
edge(b,d). 
edge(c,d).

path_disc(X,Y) :-
  edge(X,Z),
  not path_disc(Z,Y).
\end{lstlisting}
\end{minipage}
\vline depth 15ex height 15ex
\hfill
\begin{minipage}{0.36\linewidth}
\begin{lstlisting}
%%% Analysis wo. entry point %%%
:- true pred edge(_A,_B)
  : ( term(_A), term(_B) )
  => ( rt11(_A), rt10(_B) ).
:- true pred path_disc(X,Y)
  : ( term(X), term(Y) )
  => ( rt11(X), term(Y) ).

path_disc(X,Y) :-
  true((term(X),term(Y),term(Z))),
  edge(X,Z),
  true((rt11(X),term(Y),rt10(Z))),
  not_path_disc(Z,Y),
  true((rt11(X),term(Y),rt10(Z))).

:- regtype rt11/1. 
   rt11(a). rt11(b). rt11(c).
:- regtype rt10/1.
   rt10(b). rt10(c). rt10(d).
\end{lstlisting}
\end{minipage}
\vline depth 15ex height 15ex
\hfill
\begin{minipage}{0.34\linewidth}
\begin{lstlisting}
%%% Analysis with entry point %%%
% ?- initial_node(X),path_disc(X,Y)
:- true pred path_disc(X,Y)
  : ( rt19(X), term(Y) )
  => ( rt0(X), term(Y) ).

path_disc(X,Y) :-
  true((rt19(X),term(Y),term(Z))),
  edge(X,Z),
  true((rt0(X),term(Y),rt5(Z))),
  not path_disc(Z,Y),
  true((rt0(X),term(Y),rt5(Z))).

:- regtype rt0/1. rt0(a).
:- regtype rt19/1. rt19(a). rt19(d).
:- regtype rt5/1. rt5(b). rt5(c).
\end{lstlisting}
\end{minipage}
\vspace*{-2mm}
\caption{\emph{\eterms} analysis with/without entry point for the program \code|path_disc/2|.}
\label{code:pdisc}
\vspace*{-1em}
\end{figure}

Consider \code{path_disc/2} in Fig.~\ref{code:pdisc} (left), which
determines whether two nodes are disconnected by paths longer than one
edge.
Fig.~\ref{code:pdisc} (center) shows the analysis performed without
query information, i.e., all predicates are analyzed with top-most
calling patterns. The true assertions inferred by the analysis
over-approximate both calls and success patterns and record the
abstract states at each program point. This is not sufficient to
filter the OLON rule: the calling pattern \code{term(X)} intersects
with any inferred type, and the success goal for \code{not path_disc/2} %
leaves \code{Y} over-approximated as \code{term(Y)}.  This loss of
precision is caused by the \code{forall} in the dual of
\code{path_disc/2}.
In contrast, Fig.~\ref{code:pdisc} (right) shows that analyzing from
\code{?- initial_node(X), path_disc(X,Y).}, yields more precise
results.
The possible calls to \code{path_disc/2} are restricted to \code{a}
and \code{d}. Although a call with \code{X=d} may arise during
analysis, it cannot succeed because after calling \code{edge/2}, the
type of \code{X} is refined to \code{a}, since \code{edge(d,_)} 
succeeds.
Thus, the OLON loop can be omitted from inconsistency
checking, since the regular types of \code{X} and \code{Z} are
disjoint, i.e., \code{[a,d]} and \code{[b,c]}, respectively.

\vspace*{-2mm}
\section{Related work}
\label{sec:related-work}
Top-down approaches for abstract interpretation were first used in
analyzers such as MA3 and Ms~\cite{pracabsin}, and matured in
the \plai analyzer~\cite{mcctr-fixpt,ai-jlp},
which follows the tree abstraction of~\cite{bruy91}, but
implemented with an efficient fixpoint algorithm.
%
PLAI was later applied to the analysis of CLP/CHCs in~\cite{anconsall-acm}
and %
imperative programs 
\cite{anal-peval-horn-verif-2021-tplp,decomp-oo-prolog-lopstr07}.
In the context of \asp different approaches have been proposed for %
compilation, static analysis, and debugging. %
In~\cite{Costantini2005SETPT}, %
a static analysis 
methodology was proposed for analyzing an \asp program's cycle structure and
deriving consistency guarantees without computing the answer sets
directly. The approach identifies cycles and their interdependencies
and constructs a Cycle Graph, where nodes represent cycles and edges
represent connections (handles) between them. By examining this graph,
one can determine whether a program admits stable models as well as reason
about some properties.
Similarly, in~\cite{Li2021DiscASPAG}, \asp programs were represented as
Cycle Graphs and partial answer sets obtained from the resulting
structures.
Denecker et al.~\cite{Denecker2012} proposed the use of Approximation Fixpoint Theory
to define the stable and well-founded semantics of logic programs, in particular in the context of \asp. 
They %
provided %
non-monotonic semantics by means of an approximator operator. 
While a theoretical framework, %
and centered around classical ASP, 
it may be interesting to explore if similar ideas could be
useful %
in the context of s(CASP).
It has also been proposed to abstract ASP rules to sound
over-approximations of a program’s answer sets by reducing the
domain~\cite{SARIBATUR_EITER_2021}. These abstractions enable
approximate reasoning at lower cost and support unsatisfiability
explanation by identifying relevant program parts. While closely
related to Abstract Interpretation ideas, this work focuses on
over-approximating answer sets for reasoning and debugging within the
ASP context, rather than capturing more general program properties.
Finally, in~\cite{schwartz2015non} a non-monotonic program 
analysis technique was presented based on logical abduction. The approach bounds the
non-monotonic search using a monotone sequence of checkpoints,
ensuring overall convergence. The technique is applied to the analysis
of Horn clauses and similar ideas could be explored to increase
precision in the handling of the negated parts of programs.

\vspace*{-2mm}
\section{Conclusions}
\label{sec:conclusions}

We presented a first step toward Abstract Interpretation of
goal-directed ASP %
by adapting the \plai top-down fixpoint framework. %
Using the new Shared-Constraints abstract domain, we captured variable
dependencies from constraints and showed how compile-time information
can improve the execution of \scasp programs, including more efficient
\texttt{forall} evaluation.
We also explored preliminary applications in abstract specialization
and the elimination of spurious odd-loop checks.
Our results are encouraging in that they show that,
despite the current simplifications in handling the
dual rules, abstract interpretation can provide reusable program
insights and enable optimizations difficult to achieve
syntactically. This suggests opportunities for precision improvements
in future work.

\bibliographystyle{eptcs}
\bibliography{bib}

\end{document}